\documentclass[nofootinbib,prx,aps,onecolumn,floatfix,11pt,longbibliography,superscriptaddress,notitlepage]{revtex4-1}
\usepackage[utf8]{inputenc}

\usepackage{hyperref}
\usepackage{xcolor}
\usepackage{graphicx} 

\usepackage{amsthm,amsmath,amssymb,amsfonts,bbm}
\usepackage{braket}
\usepackage{bbold}

\usepackage{pythonhighlight}

\usepackage[english]{babel}
\usepackage[T1]{fontenc}
\usepackage{hyperref}

\hypersetup{
    colorlinks=true,
    linkcolor=blue,
    filecolor=magenta,      
    urlcolor=cyan,
    citecolor={blue},
    }

\usepackage{letltxmacro}
\LetLtxMacro{\ORIGselectlanguage}{\selectlanguage}
\makeatletter
\DeclareRobustCommand{\selectlanguage}[1]{%
  \@ifundefined{alias@\string#1}
    {\ORIGselectlanguage{#1}}
    {\begingroup\edef\x{\endgroup
       \noexpand\ORIGselectlanguage{\@nameuse{alias@#1}}}\x}%
}
\newcommand{\definelanguagealias}[2]{%
  \@namedef{alias@#1}{#2}%
}

\makeatother

\definelanguagealias{en}{english}
\definelanguagealias{EN}{english}
\definelanguagealias{eng}{english}
\definelanguagealias{de}{ngerman}

\begin{document}

\title{Theory of multi-qubit superradiance in a waveguide in the presence of finite 
delay times}
\author{Sofia Arranz Regidor} 
\affiliation{Department of Physics, Engineering Physics and Astronomy, Queen's University, Kingston ON K7L 3N6, Canada}

\author{Franco Nori}
\affiliation{Quantum Computing Center, RIKEN, Wakoshi, Saitama, 351-0198, Japan}
\affiliation{Physics Department, The University of Michigan, Ann Arbor, Michigan 48109-1040, USA}

\author{Stephen Hughes} 
\affiliation{Department of Physics, Engineering Physics and Astronomy, Queen's University, Kingston ON K7L 3N6, Canada}
\date{\today}

\def\bra#1{\mathinner{\langle{#1}|}}
\def\ket#1{\mathinner{|{#1}\rangle}}
\def\braket#1{\mathinner{\langle{#1}\rangle}}
\def\Bra#1{\left\langle#1\right|}
\def\Ket#1{\left|#1\right\rangle}

\begin{abstract}
We study the quantum dynamics of multiple
two-level atoms (qubits) in a waveguide quantum electrodynamics system, with a focus
on modified superradiance effects between two or four atoms with finite delay times. Using a numerically exact matrix product approach, we explore both Markovian and non-Markovian regimes, and highlight the
significant influence of time-delayed feedback effects and the clear breakdown of 
assuming instantaneous coupling dynamics.
We first show a system composed of two spatially separated qubits, prepared in a doubly excited state (both fully excited), and provide a comprehensive study of how delayed feedback influences the collective system decay rates,
as well as the quantum correlations between waveguide photons, atoms, and between atom and photons.
The system is then extended to include two additional qubits located next to the initial ones (four qubits in total), and we demonstrate, by manipulating the initial excitations and the time-delay effects, how long-term quantum correlations and light-matter entangled states can be established.

\end{abstract}

\maketitle

\section{Introduction}

Collective effects of radiatively coupled atoms and photon emitters~\cite{Ficek2002} are important for understanding fundamental aspects of light-matter interactions beyond a single-dipole decay, and have applications for producing entangled states, which is important for quantum technologies and quantum information processing.
Manifestations of collective effects include the formation of Dicke states~\cite{PhysRev.93.99,PhysRevLett.101.153601,PhysRevA.98.063815}, F\"orster coupling,  and fluorescence resonance energy transfer~\cite{Frster1948,Ravets2014,Andrews2010}.
One of the most controlled environments for 
photon-coupled two level systems (TLSs) is through waveguide modes, in an architecture known as {\it waveguide quantum electrodynamics} (waveguide QED)~\cite{PhysRevA.76.062709,PhysRevLett.98.153003,Witthaut_2010,PhysRevLett.106.053601,PhysRevLett.113.263604,Calaj2016,Hughes2004,Zheng2010,PhysRevLett.120.140404,Kannan2020,PhysRevLett.103.147003,PhysRevA.82.052509,Wilson2011,RevModPhys.84.1,PhysRevA.87.043804,GU20171}.
Waveguide QED systems offer spatial and temporal control over the
inter-atom coupling rates, and give rise to interesting quantum optical effects, such as enhanced spontaneous emission (SE)~\cite{PhysRev.69.37,PhysRevA.97.043801,PhysRevB.75.205437},  including superradiance and directional emission control~\cite{Mitsch2014,Sllner2015,doi:10.1126/science.aaa9519,Bliokh2019}. 

Many quantum optics models currently used to solve these many-body waveguide systems rely on the Markov approximation~\cite{PhysRevResearch.6.L032017,Chang_2012,Mirhosseini2019,PhysRevA.99.063829,RevModPhys.95.015002} and/or assume a weak excitation regime~\cite{PhysRevLett.98.083603,PhysRevA.101.063814}, thus often neglecting nonlinear interactions and atomic population effects~\cite{PhysRevA.82.033804,PhysRevResearch.2.043014,PhysRevA.95.053821}, as well as photon-matter entanglement (Markov regime). 
There are several well-established approaches for accurately modeling few-photon transport in waveguide QED, including scattering solutions and input-output theories that go beyond the weak excitation approximation (WEA)~\cite{PhysRevA.31.3761,Rephaeli2012FewPhotonSC,PhysRevA.82.063821,Roulet_2016,PhysRevA.102.053702,PhysRevA.102.013727,PhysRevLett.98.083603,PhysRevLett.101.100501,PhysRevA.78.063827,PhysRevA.80.062109,PhysRevA.81.042304}, but the dynamical observables can be difficult to extract directly, and often the Markov approximation is used to simplify calculations~\cite{Caneva2015}. However, the neglect of time-retardation effects cannot capture time-delayed feedback effects~\cite{ZHANG20171}, 
where more powerful techniques are needed, such as matrix product states~\cite{PhysRevResearch.3.023030}, space-discretized quantum trajectories~\cite{PhysRevA.106.013714}, or fully analytical derivations of the differential equations for simpler systems~\cite{PhysRevA.106.023708}. 

While non-Markovian delay effects are often neglected for qubits in waveguides (instantaneous coupling), especially when not invoking a WEA, it has been shown that \textit{delayed feedback effects} can greatly influence the TLS dynamics as one increases the distance between atoms~\cite{PhysRevLett.116.093601,PhysRevResearch.3.023030,PhysRevA.101.023807,PhysRevResearch.2.013238}. This is especially noticeable when considering nonlinearities and dealing with fully quantized input fields~\cite{sofia2024letter,sofia2024}. Modeling such effects is computationally challenging since, as the distance between the TLSs grows, so does the Hilbert space, and one has to treat the photons exactly, i.e., as part of the entire waveguide QED system. 

The initial state in which correlated atoms are prepared will also directly affect their behavior, e.g., it is known that a pair of symmetrically entangled TLSs ({\it superradiant state})~\cite{PhysRev.93.99,Gross1982,Anatolii,PhysRevA.96.023863,PhysRevB.94.224510} will cause a faster SE decay of the atoms and, it has been shown that finite distances between them can further enhance the atom SE decay, where a constructive relative phase between the atoms can lead to a collective emission higher than the usual superradiant one; this is the so-called ``superduperradiance''~\cite{PhysRevLett.124.043603}.
Thus, finite delay effects can modify the usual understanding of Dicke states, where non-Markovian feedback effects are not taken into account, and $N$ atoms starting in a symmetrical state have an initial 
{\it temporal burst}, enhancing their collective decay. In the Markovian case, a fully excited $N$-qubit system (i.e., starting with all atoms excited), 
will have a collective decay rate of~\cite{PhysRev.93.99,Gross1982}
\begin{equation}
    W_N= N\gamma,
    \label{Nexc}
\end{equation}
while a half excited symmetrical initial state, where $N$ is an even number, will have a decay rate of
\begin{equation}
    W_{N/2}= \frac{N}{2} \left(\frac{N}{2} + 1\right) \gamma,
    \label{N/2exc}
\end{equation}
and in the limit where there is only one initial excitation, the excitation will decay with 
\begin{equation}
    W_1=N \gamma.
    \label{1exc}
\end{equation}

\begin{figure}[hb]
\centering
\includegraphics[width=0.49\columnwidth]{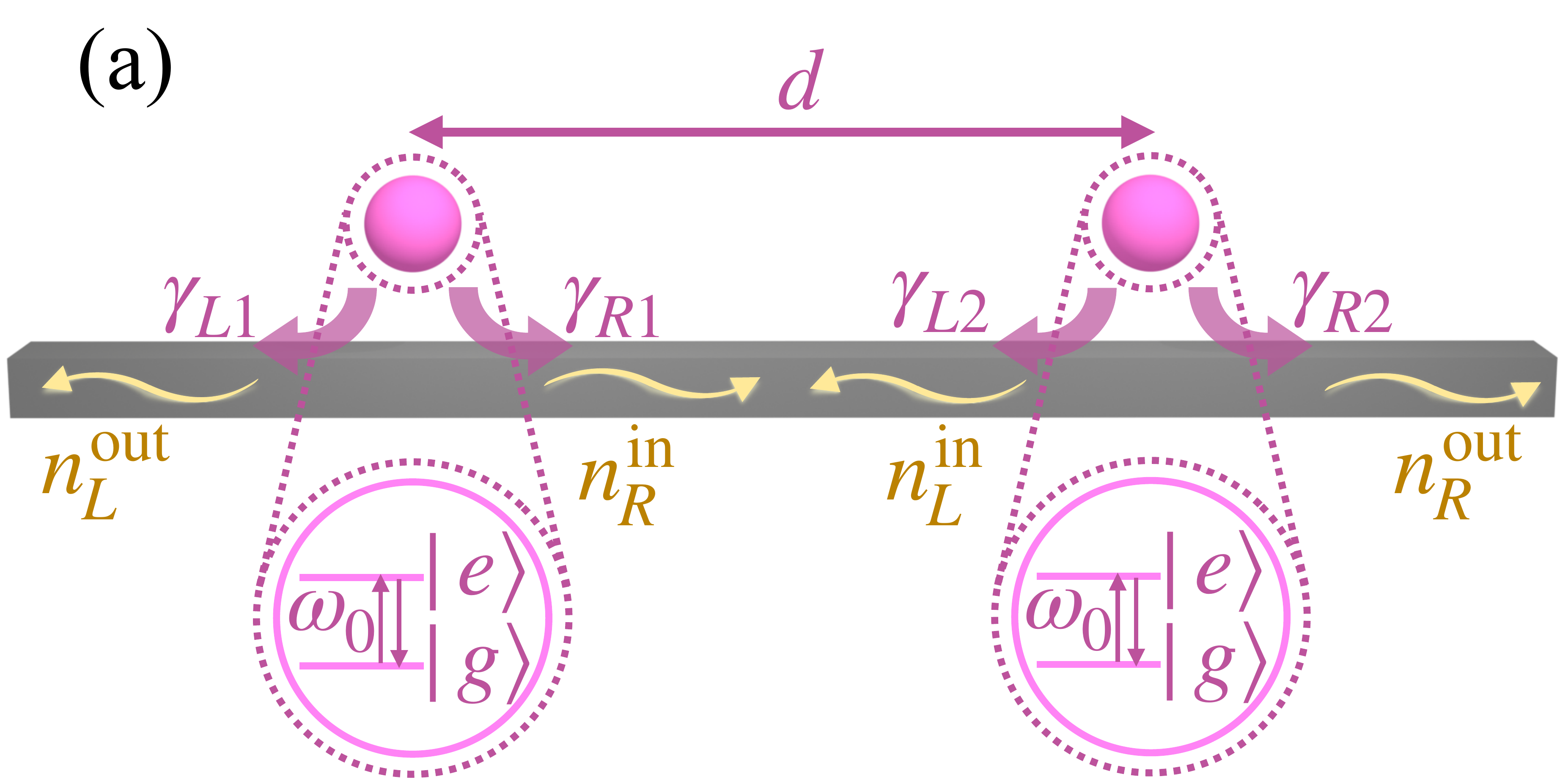}
\includegraphics[width=0.49\columnwidth]{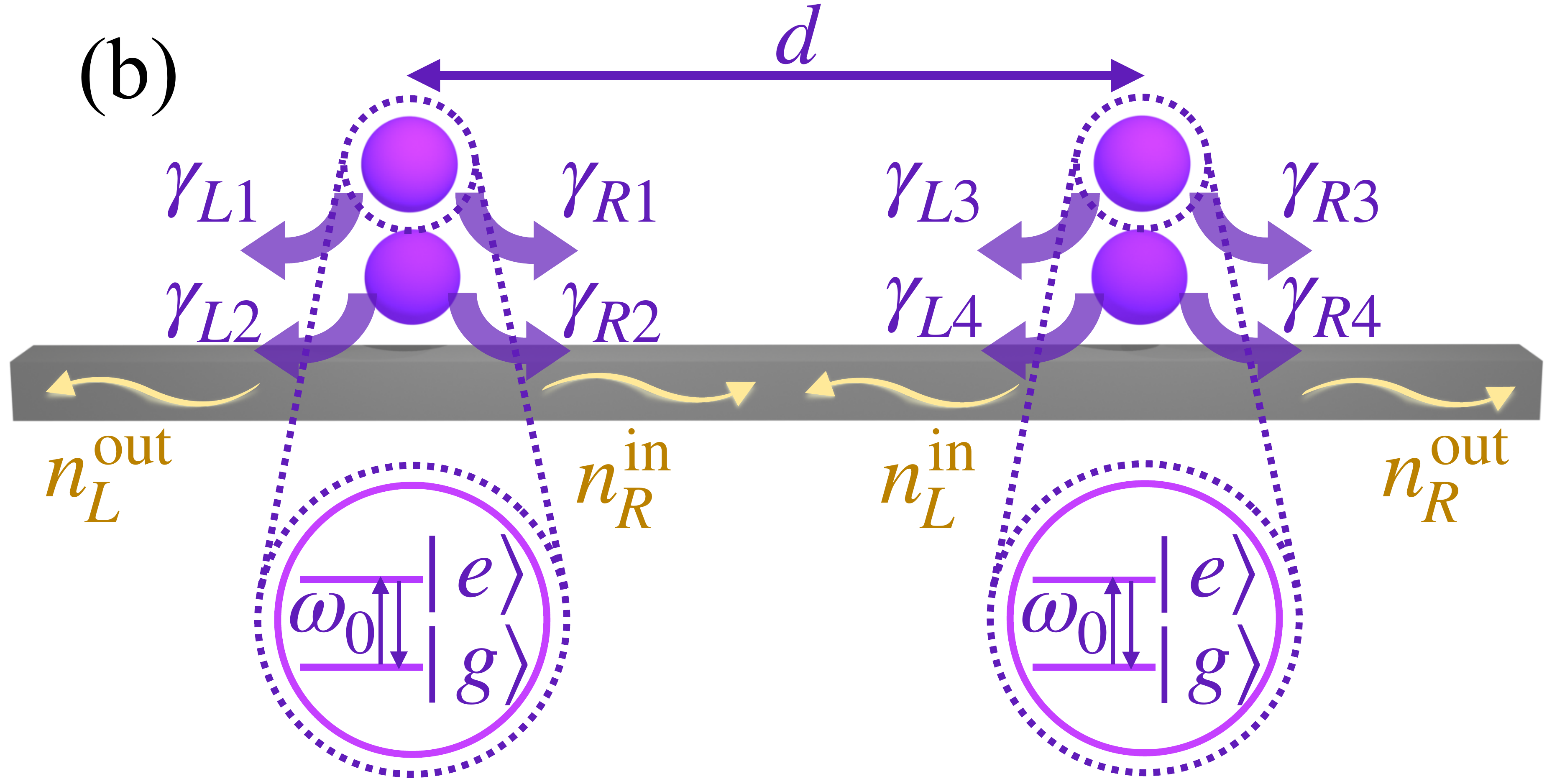}
\caption{(a) Schematic of two artificial atoms (treated as TLSs) in an infinite (open) waveguide, separated by a finite distance $d$. (b) Similar system to (a) but now with four TLSs, two in each of the groups. 
}
\label{fig:schem}
\end{figure}

\subsection{Finite delay effects in waveguide QED}

In this work, \textit{we investigate waveguide QED finite delay effects (yielding non-Markovian effects as seen by the atoms)} with 
a pair of TLSs prepared in an initial state with two quantum excitations (thus, beyond a simple linear response), and \textit{explore the role of finite delays as well as the effects of adding extra TLSs next to the initial pair}. To solve this complex many-body problem, we use a tensor network approach (specifically, using matrix product states or MPS) \cite{PhysRevResearch.3.023030,orus_practical_2014,yang_matrix_2018,vanderstraeten_tensor_2017,Naumann2017,Droenner2019,PhysRevX.14.031043}, which allows us to decompose a large Hilbert space in smaller subspaces and solve the problem in a numerically exact and efficient way. Therefore, we can fully include finite delay and feedback effects, nonlinearities, and a higher number of quantized excitations, and easily extract useful information and observables both for the atoms and fields (quantized waveguide photons). 

Although the collective response of identical atoms is a well studied problem, it has been predominantly studied in the Markovian regime, thus ignoring finite-delay feedback effects. These Markovian approaches fail when delay times become comparable to the lifetime of the TLSs (and often delay times with just a fraction of this lifetime can induce fundamentally new correlation effects~\cite{PhysRevA.106.013714}), which can have a pronounced effect on the quantum dynamics and quantum correlation functions. Fully non-Markovian dynamics have been studied mostly in the long-time limit~\cite{PhysRevA.102.043718,Dinc2019} or considering a single excitation~\cite{PhysRevLett.124.043603} (in the WEA). However,  recent work~\cite{PhysRevResearch.6.023213} has exploited  Laplace transform solutions to analytically solve the time dynamics of two atoms initially excited, separated by a distance $d=\tau v_g$, where $\tau$ is the delay time between them, and $v_g$ is the group velocity of the electromagnetic field; the authors show the probability of having one excitation versus time, as well as the concurrence (atom-atom entanglement) and correlation between atomic dipoles. Although the feedback effects are clearly apparent in these three observables, the work also 
assumes that the probability of having two excitations is independent of the delay time. However, in our work, \textit{we show how the two-atomic excitation probability does depend on the finite feedback effects}, and we 
significantly extend the study of such a system to explore other observables, thus gaining a deeper understanding of this problem. In addition, we also explore the role of adding in additional atoms,
as discussed below. 

\subsection{Two TLSs separated a distance $d$}

In our study, we will first consider two TLSs separated by a finite distance $d$, shown in Fig.~\ref{fig:schem}(a), where the two atoms start in a two-excitation superradiant state, with a destructive phase between both atoms; this case is intrinsically nonlinear. To highlight the physics of the finite-delay effects, we solve the problem in two regimes: (i) in the Markovian limit (instantaneous coupling regime), and (ii) in the non-Markovian regime (with delayed feedback effects fully included), and investigate how the distance between the atoms affects the quantum dynamics. 

We investigate not only the atom and field dynamics but also the correlations between them, calculating correlations between the two TLSs, field (photon) correlations and light-matter correlations that give us a deeper understanding of the dynamics, showing signatures of long-lasting light-matter entangled states. We also explore how finite delays affect the atomic excitations, modifying the doubly excited collective decay, contrary to previous findings in the literature, where the decay of a doubly excited state was independent of the distance between TLSs~\cite{PhysRevResearch.6.023213}. 

\subsection{Multi-atom waveguide QED results}

In addition, we also study multi-atom wQED results when the system is extended to include additional atoms located close to each of the initial TLS pair, as shown in Fig.~\ref{fig:schem}(b), where we now have additional degrees of freedom when choosing an initial state. We will demonstrate significant deviations from the usual understanding of collective Dicke states in a Markovian regime, such as population trapping of the double-exited state and photon trapping in the waveguide region between the TLS pair, thus offering a much richer coupling regime with the non-Markovian coupling and control. 
We also show how adding one TLS next to each of the initial atoms affects the collective emission of the system. Here, we can either add them as initial unexcited atoms and thus start the system in a product state, or, alternatively, engineer new initial superradiant states, e.g., starting each pair in an entangled state with a shared excitation, that may enhance the collective SE decay rates, or maximize the field entanglement. Moreover, we calculate various other useful observables such as the transmitted or reflected photon fluxes and light-matter quantum correlations, including emitted spectra, to better understand the unique nature of these complex quantum systems. 

The rest of our paper is organized as follows. In Sec.~\ref{sec:theory}, we introduce our MPS theory, and the theoretical approach chosen to solve these waveguide QED problems, with the solution for two TLSs in a waveguide in Sec.~\ref{subsec:theory2}, and the extension to four TLSs (two pairs, spatially separated) in Sec.~\ref{subsec:theory4}. Section~\ref{sec:results2} studies the time dynamical results for two TLSs, including a study of the delay dependence on the probability of atomic excitations in Sec.~\ref{subsec:P2}.
In Sec.~\ref{subsec:further_an}, we show a further study of the single atomic excitation as a function of the delay times, and investigate the atom-atom correlations, and the role of the phase between atoms in Sec.~\ref{subsec:phase_dep}. Additional quantum observables are calculated in Sec.~\ref{subsec:more_obs}, providing a comprehensive understanding of the system. Afterwards, in Sec.~\ref{sec:results4}, we analyze the influence of adding two new TLSs next to the previous ones, where we now have more degrees of freedom starting with an initially doubly-excited state. Here, we study again the collective decay rates and photon fluxes in the non-Markovian regime, showing long-lasting light-matter entangled states. Finally, in Sec.~\ref{sec:conclusions} we present our closing remarks and conclusions.

\section{Theoretical Description}
\label{sec:theory}

\subsection{A pair of two level systems in an infinite waveguide}
\label{subsec:theory2}

A system consisting of two TLSs in an infinite waveguide [Fig.~\ref{fig:schem}(a)], can be represented by the following Hamiltonian~\cite{PhysRevResearch.3.023030}
(units of $\hbar=1$),
\begin{equation}
     H = \sum_{n=1,2} H_{\rm TLS}^{(n)} + H_{\rm W} + H_{\rm I},
    \label{hamil2tls}
\end{equation}
where the first term represents the TLSs,
\begin{equation}
    H_{\rm TLS}^{(n=1,2)}= \omega_n \sigma_n^+ \sigma_n^-
    \label{TwoTLS_H},
\end{equation}
with $\omega_n$ the transition frequency, and $\sigma_n^+$, $\sigma_n^-$ the relevant Pauli operators for TLS $n$.
The waveguide term is
\begin{equation}
    H_{\rm W} = \sum_{\alpha=L,R} \int_{\mathcal{B}} d\omega  \omega b_\alpha^\dagger (\omega)b_\alpha(\omega),
\end{equation}
where $b_\alpha^\dagger (\omega)$ and $b_\alpha(\omega)$ are the creation and annihilation bosonic operators, respectively, for both right- and left-moving photons, and $\mathcal{B}$ is the relevant bandwidth of interest around the resonance frequency $\omega_0$, in the rotating-wave approximation. 
Finally, the last term represents the interaction
between the photons and qubits, 
\begin{equation}
\begin{split}
    H_{\rm I} &=  \frac{1}{\sqrt{2\pi}}\int_{\mathcal{B}} d\omega \Bigg \{ \sum_{n=1,2}
    \left( \sqrt{\gamma_{L_n}} e^{i\omega x_n/c}\, b_L(\omega)\sigma^+_n \right .
     \left . +  \sqrt{\gamma_{R_n}} e^{-i\omega x_n/c} b_R(\omega)\sigma^+_n 
     \right) + \rm H.c. \Bigg \}, 
     \label{H_I}
\end{split}
\end{equation} 
where $x_1$ and $x_2$ are the locations of each TLS in the waveguide, $d=x_2-x_1$ is the distance between them, and $\tau=d/c$ is the \textit{delay time} that it takes for a photon to travel from one TLS to the other.
Here, we have assumed that the coupling frequency is close to the TLSs frequency, and that the decay rates $\gamma_{L}$ and $\gamma_{R}$ to the left and right, respectively, do not depend on frequency.
We also transform to the interaction picture with a rotating frame in terms of $\omega_0$.
Next, we transform the Hamiltonian to the interaction picture with respect to the free dynamics, and then transform it to the time domain, where we change accordingly the boson operators ~\cite{PhysRevResearch.3.023030}, 
and use MPS to solve the problem, without performing any bath approximations. 

Matrix product states is an approach based on tensor network methods that uses Schmidt decompositions or singular value decompositions to decompose the large Hilbert space of the entire system, including both the TLSs and the waveguide, into smaller subspaces. The significant advantage of this approach is that, once we have these smaller subspaces, we can apply the time evolution operator only in the relevant parts of the system at each time step~\cite{vanderstraeten_tensor_2017,mcculloch_density-matrix_2007,schollwock_density-matrix_2011}.

To proceed, we discretize the waveguide in time into the so-called ``time bins'' and write the discretized Hamiltonian in time~\cite{Pichler11362,PhysRevResearch.3.023030,droenner_out--equilibrium_2019}.
Thus, we can write the time evolution operator for each time step to evolve the whole system, 
\begin{equation}
    U(t_{k+1},t_k) =  \exp{ \left( -i \int_{t_k}^{t_{k+1}} dt' H(t')\right)}, 
\end{equation}
which in this case can be written as
\begin{equation}
    \begin{split}
    U(t_{k+1},t_k) &=  \exp \Bigg \{ 
    -i \left[ \left( \sqrt{\gamma_{L1}} \Delta B _L (t_k) + \sqrt{\gamma_{R1}} \Delta B_R (t_{k-l}) e^{i\phi}\right)\sigma^+_1 +\rm H.c. \right] \\
    &-i \left[ \left( \sqrt{\gamma_{L2}} \Delta B_L(t_{k-l})e^{i\phi} + \sqrt{\gamma_{R2}} \Delta B_R(t_k) \right)\sigma^+_2 + \rm H.c. \right] \Bigg \} , 
    \end{split}
    \label{u2tls}
\end{equation}
where $\Delta B_{R/L} ^{(\dagger)}  = \int_{t_k}^{t_{k+1}} dt' b_{R/L}^{(\dagger)}(t')$ are the noise matrix product operators (MPOs) that create/annihilate a photon in a 
{\it time bin} with a commutation relation 
\begin{equation}
     \left[ \Delta B_{R/L}(t_k), \Delta B_{R/L}^\dagger(t_{k'}) \right] = \Delta t \delta_{k,k'} \delta_{R,L},
\end{equation}
and $l=\tau/\Delta t$ represents the number of time bins between the two TLSs~\cite{PhysRevResearch.3.023030}.
In the Markovian limit, $\tau=0$, and everything is assumed to happen at the same time step (instantaneous coupling regime).
The propagation phase, $\phi$, between the two TLSs, in the interaction picture, is defined from
\begin{equation}
    \phi=-\omega_0 \tau, 
    \label{eq:phase}
\end{equation}

Unless there is an external pump field, the number of excitations in the system should be conserved (at least without the rotating-wave approximation), hence, a conservation rule must be fulfilled. For example, considering two TLSs in the Markovian regime, the total number of excitations,
$N_{\rm total} (t)$, for all time, $t$,  is
\begin{equation}
\begin{split}
    N_{\rm total} (t) &= n_{\rm TLS1} (t) + n_{\rm TLS2} (t) + \int_0^t \left( n_{R}^{\rm out}(t') +n_{L}^{\rm out}(t') \right) dt' \\
    &= n_{\rm TLS1} (t) + n_{\rm TLS2} (t) + N^{\rm out}_R (t) + N^{\rm out}_L (t),
    \end{split}
 \label{Ntotal}   
\end{equation}
where $n_{\rm TLS1}= \braket{\sigma^+_1 \sigma^-_1}$ and $n_{\rm TLS2}=\braket{\sigma^+_2 \sigma^-_2}$ are the emitter populations of the TLSs, and $n^{\rm out}_{R/L}(t)= \braket{b_{R/L}^\dagger (t) b_{R/L}(t)}$ and $N^{\rm out}_{R/L}$ are the right/left integrated fluxes (proportional to the photon populations). For example, if we start with both TLSs excited ($\ket{\phi_0}=\ket{ee}$), then $N_{\rm total} (t) = 2$ at all times.

In the non-Markovian regime, the conservation rule will have two extra terms, since we now have to consider the possibility of having {\it photons within the feedback loop}, i.e., in the waveguide space between the qubits, so we have
\begin{equation}
\begin{split}
    N_{\rm total}(t) &= n_{\rm TLS1}(t) + n_{\rm TLS2}(t) + \int_{0}^t (n_R^{\rm out} (t') + n_L^{\rm out}(t')) dt' + \int_{t - \tau}^t (n_R^{\rm in} (t') + n_L^{\rm in}(t')) dt' \\
    &= n_{\rm TLS1}(t) + n_{\rm TLS2}(t) + N^{\rm out}_R(t) + N^{\rm out}_L(t) + N^{\rm in}_R(t) + N^{\rm in}_L(t),
\end{split}
\label{Ntotaltau}
\end{equation}
where the new integral terms account for the photons within the feedback loop ($N^{\rm in}_{R/L}$). Once again, if we start with both fully excited TLSs, $N_{\rm total} (t) = 2$ at all times.

Next, we define the time evolution of the probability of having zero [$P^{(0)}(t)$], one [$P^{(1)}(t)$] or two [$P^{(2)}(t)$] atomic excitations in the waveguide system, where $P^{(0)}(t)$ corresponds to having both TLSs in the ground state, $P^{(1)}(t)$ is the probability of having one of them excited while the other one stays unpopulated, which in this case will be a combination of the probability of having either the left or right TLS populated, and $P^{(2)}(t)$ corresponds to having both TLSs populated:
\begin{equation}
    P^{(0)} (t) = \braket{\sigma^-_1 \sigma^+_1 \sigma^-_2 \sigma^+_2}(t),
\end{equation}
\begin{equation}
    P^{(1)} (t) = \braket{\sigma^+_1 \sigma^-_1 \sigma^-_2 \sigma^+_2} (t)+ 
    \braket{\sigma^-_1 \sigma^+_1 \sigma^+_2 \sigma^-_2}(t),
\end{equation}
\begin{equation}
    P^{(2)} (t) = \braket{\sigma^+_1 \sigma^-_1 \sigma^+_2 \sigma^-_2}(t).
\end{equation}

We can again write the conservation rule now in terms of these probabilities for the case with two excitations,
%
\begin{equation}
\begin{split}
    N_{\rm total}(t) &= P^{(1)}(t) + 2P^{(2)}(t) + N^{\rm out}_R(t) + N^{\rm out}_L(t) + N^{\rm in}_R(t) + N^{\rm in}_L(t) = 2,
    \label{cons_rule}
\end{split}
\end{equation}
considering that in the Markovian limit $N^{\rm in}_{R/L}(t) = 0$.

\subsection{Two pairs of qubits (four TLSs) in an infinite waveguide}
\label{subsec:theory4}

Next we consider two additional TLSs, next to the initial TLSs that we previously had, thus the system is now formed by two TLSs at each side [Fig.~\ref{fig:schem} (b)], with four in total.
The complete system Hamiltonian is now
\begin{equation}
     H = \sum_{n=1,2,3,4} H_{\rm TLS}^{(n)} + H_{\rm W} + H_{\rm I},
    \label{hamil4tls}
\end{equation}
with the interaction term,
\begin{equation}
\begin{split}
    H_{\rm I} &=  \frac{1}{\sqrt{2\pi}}\int_{\mathcal{B}} d\omega \Bigg \{ \sum_{n=1,2}
    \left( \sqrt{\gamma_{L_n}} e^{i\omega x_L/c}\, b_L(\omega)\sigma^+_n \right .
     \left . +  \sqrt{\gamma_{R_n}} e^{-i\omega x_L/c} b_R(\omega)\sigma^+_n 
     \right) + \rm H.c. \Bigg \} \\
     &+\frac{1}{\sqrt{2\pi}}\int_{\mathcal{B}} d\omega \Bigg \{ \sum_{n=3,4}
    \left( \sqrt{\gamma_{L_n}} e^{i\omega x_R/c}\, b_L(\omega)\sigma^+_n \right .
     \left . +  \sqrt{\gamma_{R_n}} e^{-i\omega x_R/c} b_R(\omega)\sigma^+_n 
     \right) + \rm H.c. \Bigg \}
     , 
     \label{H_I2}
\end{split}
\end{equation} 
where $x_L$ represents the location of the TLSs situated on the left, and $x_R$ for right location, and $d=x_R-x_L$ is the distance between the pairs (as before).

The modified time evolution operator is 
\begin{equation}
    \begin{split}
    U(t_{k+1},t_k) &=  \exp \Bigg \{ 
    -i \left[ \left( \sqrt{\gamma_{L1}} \Delta B _L (t_k) + \sqrt{\gamma_{R1}} \Delta B_R (t_{k-l}) e^{i\phi}\right)\sigma^+_1 +\rm H.c. \right] \\
    &-i \left[ \left( \sqrt{\gamma_{L2}} \Delta B _L (t_k) + \sqrt{\gamma_{R2}} \Delta B_R (t_{k-l}) e^{i\phi}\right)\sigma^+_2 +\rm H.c. \right] \\
    &-i \left[ \left( \sqrt{\gamma_{L3}} \Delta B_L(t_{k-l})e^{i\phi} + \sqrt{\gamma_{R3}} \Delta B_R(t_k) \right)\sigma^+_3 + \rm H.c. \right]
    \\
    &-i \left[ \left( \sqrt{\gamma_{L4}} \Delta B_L(t_{k-l})e^{i\phi} + \sqrt{\gamma_{R4}} \Delta B_R(t_k) \right)\sigma^+_4 + \rm H.c. \right]
    \Bigg \},
    \end{split}
    \label{u4tls}
\end{equation}
and in the following sections, we will consider identical atoms, with $\gamma_{Rn} = \gamma_{Ln} = \gamma/2$ (symmetric coupling), though this is not a model requirement. 

If we study the system in terms of atomic excitations, in the case of still having two total excitations, the conservation rule still follows, Eq.~\eqref{cons_rule}, but the atomic probabilities 
are now:
\begin{equation}
    P^{(0)} (t) = \braket{\sigma^-_1 \sigma^+_1 \sigma^-_2 \sigma^+_2
    \sigma^-_3 \sigma^+_3\sigma^-_4 \sigma^+_4}(t),
\end{equation}
\begin{equation}
\begin{split}
    P^{(1)} (t) &= \braket{\sigma^+_1 \sigma^-_1 \sigma^-_2 \sigma^+_2 \sigma^-_3 \sigma^+_3 \sigma^-_4 \sigma^+_4 }(t) + 
    \braket{\sigma^-_1 \sigma^+_1 \sigma^+_2 \sigma^-_2  \sigma^-_3 \sigma^+_3 \sigma^-_4 \sigma^+_4}(t) \\
    &+
    \braket{\sigma^-_1 \sigma^+_1 \sigma^-_2 \sigma^+_2  \sigma^+_3 \sigma^-_3 \sigma^-_4 \sigma^+_4}(t) +
    \braket{\sigma^-_1 \sigma^+_1 \sigma^-_2 \sigma^+_2    \sigma^-_3 \sigma^+_3 \sigma^+_4 \sigma^-_4} (t)
    ,
\end{split}
\end{equation}
\begin{equation}
\begin{split}
    P^{(2)} (t) &= \braket{\sigma^+_1 \sigma^-_1 \sigma^+_2 \sigma^-_2 \sigma^-_3 \sigma^+_3 \sigma^-_4 \sigma^+_4 }(t) + 
    \braket{\sigma^-_1 \sigma^+_1 \sigma^+_2 \sigma^-_2  \sigma^+_3 \sigma^-_3 \sigma^-_4 \sigma^+_4}(t) \\
    &+
    \braket{\sigma^-_1 \sigma^+_1 \sigma^-_2 \sigma^+_2  \sigma^+_3 \sigma^-_3 \sigma^+_4 \sigma^-_4}(t) +
    \braket{\sigma^+_1 \sigma^-_1 \sigma^-_2 \sigma^+_2    \sigma^-_3 \sigma^+_3 \sigma^+_4 \sigma^-_4}(t) \\
    &+ \braket{ \sigma^+_1 \sigma^-_1 \sigma^-_2 \sigma^+_2  \sigma^+_3 \sigma^-_3 \sigma^-_4 \sigma^+_4}(t)+
    \braket{ \sigma^-_1 \sigma^+_1 \sigma^+_2 \sigma^-_2 \sigma^-_3 \sigma^+_3 \sigma^+_4 \sigma^-_4 } (t).   
\end{split}
\end{equation}

With our MPS approach, we note that this can be easily extended to
include the addition of more pairs of atoms located next to the current ones (or at other spatial locations as well).

\section{Results for two coupled qubits with two excitations}
\label{sec:results2}

\subsection{Finite delay dependence on the probability of atomic excitations with two qubits}
\label{subsec:P2}

We start by looking at two TLSs, initially excited, separated by a distance $d$ [see Fig.~\ref{fig:schem}(a)], and study the TLS population
dynamics as well as how feedback effects can affect their decay rates.

In the Markovian regime, one has a traditional Dicke state and, as mentioned above, the collective decay rate simply depends on the number of TLSs and the initial number of initial excitations (with one excitation being inherently linear). Here we have two TLSs, hence $N=2$, and particularly in this case, if the system starts with one excitation, following Eq.~\eqref{1exc}, $W_1= 2 \gamma$; 
but if there are initially two excitations, from Eq.~\eqref{Nexc}, $W_2= 2 \gamma$ too.
This means that in the case of one excitation, the decay will be $P^{(1)}(t) = e^{-2\gamma t}$, and for two TLSs initially excited will have a collective decay $P^{(2)}(t) = e^{-2\gamma t}$.

\begin{figure}[h]
    \centering
    \includegraphics[width=\linewidth]{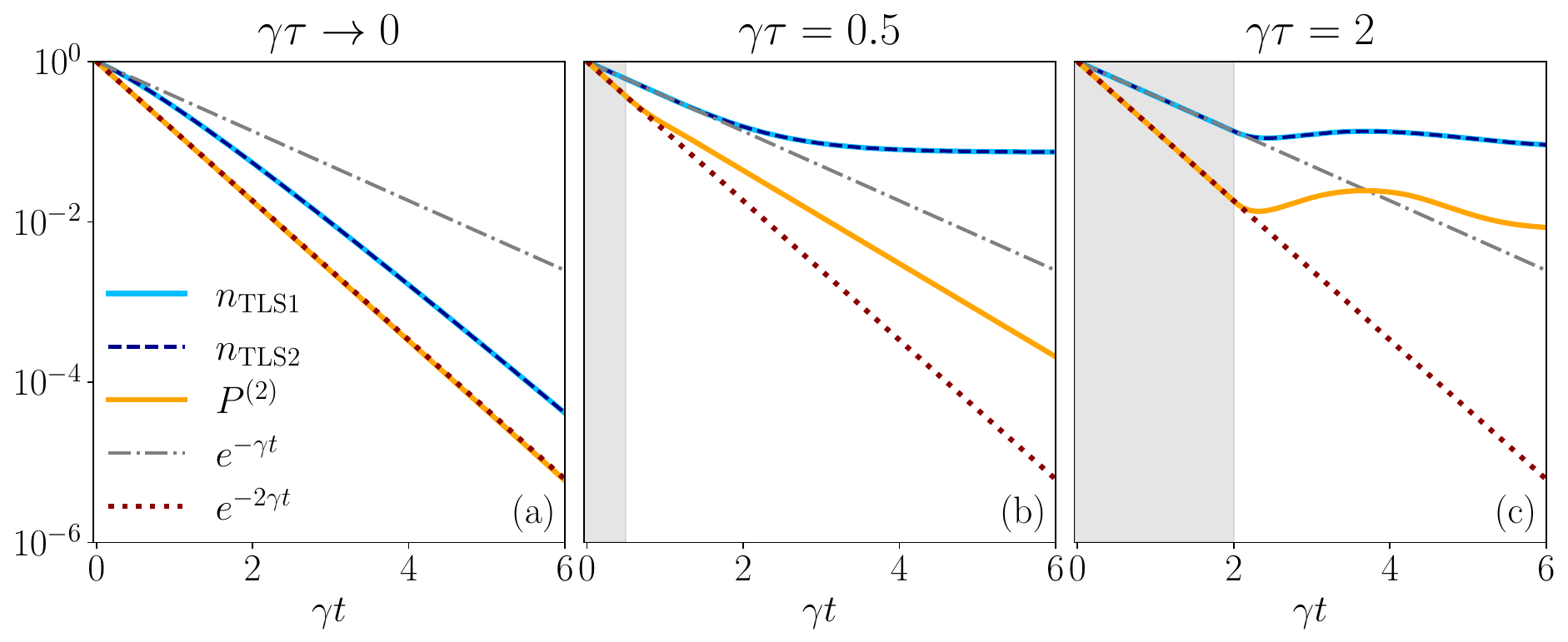}
    \caption{Emitter population dynamics of two TLSs, initially excited, solved in the Markovian regime in (a); then with a delay time of $\gamma \tau = 0.5$ in (b), and with a delay time of $\gamma \tau = 2$ in (c). Individual TLS populations are shown in blue (solid and dashed), the probability of two atomic excitations [$P^{(2)} (t)$] in orange, and the analytical exponential decays for the Markovian regime are shown in (a), for reference, with the dash-dotted grey line and dotted maroon line, respectively, for 1 and 2 initial excitations).
    The shaded grey area in (b-c) indicates the delay time, $\tau$, when the non-Markovian dynamics begins to occur.
    }
    \label{fig:log2exc}
\end{figure}

Figure~\ref{fig:log2exc} shows the population dynamics of each atom decaying from an initially doubly excited state ($\ket{\phi_0} = \ket{ee}$), and the collective doubly-excited atom decay [($P^{(2)} (t)$]. Figure~\ref{fig:log2exc}(a) presents the 
Markovian dynamics matching the analytical decay of $2 \gamma$ shown with the red chain curve. We then consider delay-induced feedback effects, for two different delay times $\gamma \tau =0.5$ and $\gamma \tau=2$ in Figs.~\ref{fig:log2exc}(b) and (c), with a phase $\phi=0$ in both cases. With a time-delayed feedback, we observe the trapping of the TLS populations, and with the longer delay time, we can also observe oscillations due to the partial re-excitation of the TLSs; this oscillation time corresponds to the photon roundtrip time, which is clearly a significant delay-induced dynamic.

We also observe how the probability of having two atomic excitations is highly affected by the feedback effects, even with the shorter delay time. This is in contrast to previous work reported in Ref.~\onlinecite{PhysRevResearch.6.023213}, which derives $P^{(2)}(t)=e^{-2 \gamma t}$, and states that the probability of having two atomic excitations is {\it independent of the delay between atoms}.
However, Fig.~\ref{fig:log2exc} clearly shows that it does depend on the delay times, and it becomes significant for larger delay times. It is important to also note that we have calculated the conservation rule,
using both the individual TLS populations [see Eq.~\ref{Ntotaltau}] and the collective atomic excitations [see Eq.~\ref{cons_rule}], in order to confirm the validity of our results.

\subsection{Further analysis of the one atomic excitation probability, $P^{(1)}$,  and atom-atom correlations}
\label{subsec:further_an}

\begin{figure}[h]
    \centering
    \includegraphics[width=0.45\linewidth]{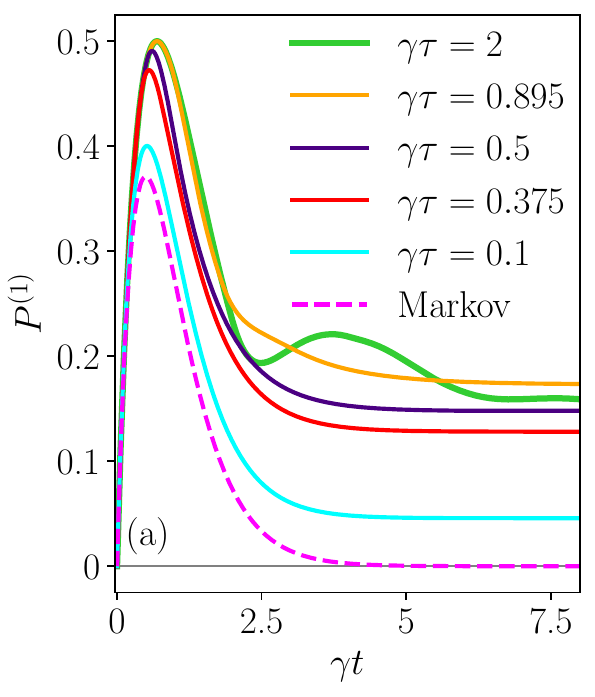}
    \includegraphics[width=0.48\linewidth]{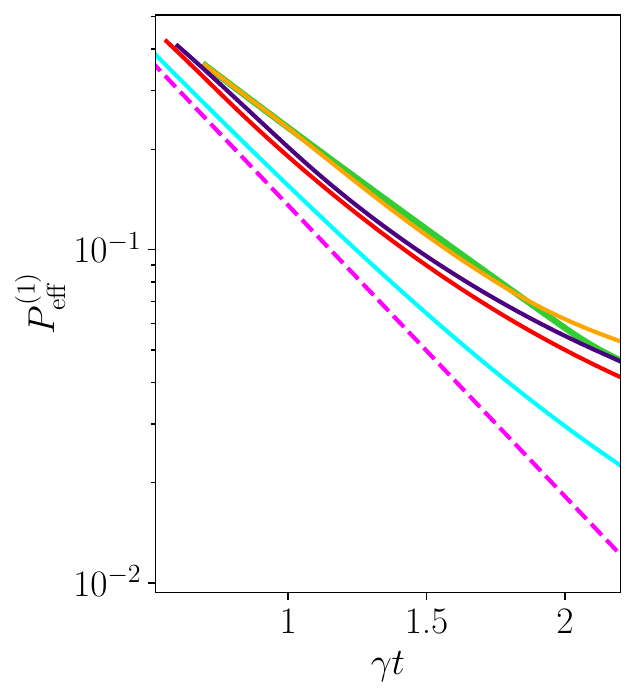}
    \caption{ Time dynamics of two atoms initially excited.
    (a) Probability of having only one of the TLSs excited, i.e. probability of one atomic excitation $P^{(1)}(t)$, for different delay times, when the system is initialized with both TLSs excited. (b) Zoom-in of the reduced probability of having only one of the TLSs excited $P^{(1)}_{\rm eff} (t)$ on a logarithmic scale, for the different delay times plotted in (a).
    }
    \label{fig:P1taus}
\end{figure}

To more fully explore the non-Markovian effects of our system, we also search for interesting delay times (or atom separations) that can lead to interesting temporal effects that qualitatively change because of finite-delay coupling regimes between the emitters. In Ref.~\cite{PhysRevResearch.6.023213}, for the case of two atoms initially excited, the authors studied two specific lengths that were cited to correspond to the largest instantaneous decay rate ($\gamma \tau =0.375$), and the maximum probability of having 1 atomic excitation in the long time limit, max. $P^{(1)} (t \to \infty)$ ($\gamma \tau =0.895$).

To have a more complete picture, we study $P^{(1)} (t)$ for a number of different delay times in Fig.~\ref{fig:P1taus}(a), including the ones mentioned above. We observe that, as stated in Ref.~\cite{PhysRevResearch.6.023213}, $\gamma \tau =0.895$ indeed gives the highest value of $P^{(1)} (t)$ at long times.  
However, it is not clear that the delay time of $\gamma \tau =0.375$ yields the largest instantaneous decay rate, since in Fig.~\ref{fig:P1taus}(a) this does not appear to be the case.

In the Markovian regime, $P^{(1)} (t)$ has a known analytical solution, $P^{(1)} (t) = 2 \gamma t e^{-2\gamma t}$~\cite{Gross1982}. 
Thus, we can define  $ P^{(1)}_{\rm eff} (t) = P^{(1)} (t)/2 \gamma t$
to extract its effective decay rate, which in this case will be $2\gamma$.
Of course, this assumes the solution remains Markovian, so the results below are for a qualitative understanding of the effective decay rates.
In Fig~\ref{fig:P1taus}(b), we show $P^{(1)}_{\rm eff} (t)$ for both the same results shown in (a), on a logarithmic scale. First, we see how, as expected, the Markovian result is a straight line with slope of $2\gamma$. Second, we observe how feedback effects create new nonlinearities and the decay rates are time dependent; however, we do not see an especially faster decay for $\gamma \tau = 0.375$, and the original derivation of this value is uncertain to us, so it is not clear whether the choice of $\gamma \tau =0.375$ corresponds to the largest instantaneous decay rate of this system or not.

\begin{figure}[h]
\centering
\includegraphics[width=0.7\columnwidth]{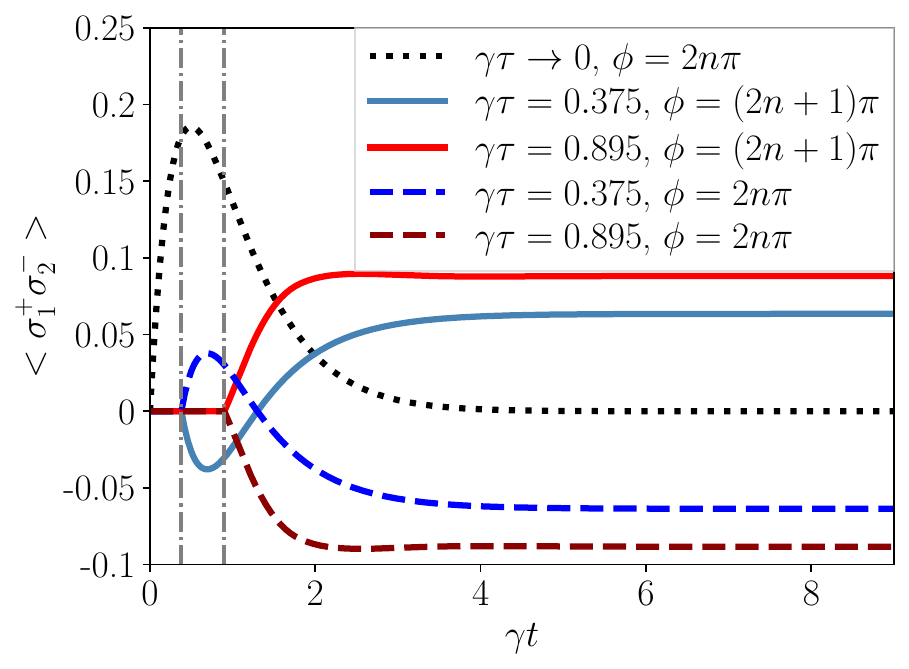}
\caption{Inter-emitter correlations calculated using MPS for two TLSs separated by $\gamma \tau = 0.375$ and $\gamma \tau = 0.895$, where we distinguish between even and odd multiples of $\pi$ for the phase, $\phi = 2n \pi$ and $\phi = (2n+1) \pi$. The Markov limit ($\gamma \tau \to 0$) is shown for comparison.
}
\label{fig:paperpop}
\end{figure}

To help study the underlying physics of these
different delays and how they affect the inter-emitter coupling, in Fig.~\ref{fig:paperpop}, we calculate the emitter-emitter correlation function, specifically $\braket{\sigma^+_1 \sigma^-_2}(t)$, in order to benchmark and extend previous atom-atom correlation results shown in Ref.~\cite{PhysRevResearch.6.023213}, shown in their figure~3(b)\footnote{We use `figure' to refer to graphs in \cite{PhysRevResearch.6.023213} to avoid confusion with our own text graphs. which are labeled with Fig.}, where results for $\gamma \tau = 0.895$ are shown. We highlight that our findings appear to significantly differ from the ones reported in~\cite{PhysRevResearch.6.023213}. First, when we look at the $\gamma \tau = 0.895$ with $\phi = \pi$ solution, which corresponds to the red solid line in Fig.~\ref{fig:paperpop}, we observe different 
time dynamics, with only positive values of the atomic correlations. We are not clear on where this discrepancy comes from, but we have double-checked our own results.  However, if we now calculate this atomic correlation function for $\gamma \tau = 0.375$ (blue solid line), the time dynamics follow a similar shape to the case reported in Ref.~\cite{PhysRevResearch.6.023213}, their figure~3(b), {\it but} with their simulated delay time of $\gamma \tau = 0.895$. In our Fig.~\ref{fig:paperpop}, when considering $\gamma \tau = 0.375$ with say $\phi=\pi$ (blue solid), the atomic correlation goes to negative values immediately after the finite delay time, and then transitions to positive values at longer times, similar to the solution presented in  figure~3(b) of ~\cite{PhysRevResearch.6.023213}, again with their longer delay time of $\gamma \tau = 0.895$,
except for an earlier start (i.e., when we first see a temporal change) in our results due to the shorter delay time. 

Additionally, we note that one has to be careful when choosing a value of $m$ for the $\phi = m \pi$ phase. Here, we find that there is a difference in the results between an even or odd value of $m$, with a flip on the sign when $m$ is even. This is shown in Fig.~\ref{fig:paperpop} in dashed curves. Furthermore, for phases $\phi = \frac{(2m + 1) \pi}{2}$, there is no emitter-emitter correlation, and $\braket{\sigma^+_1 \sigma^-_2} (t)= 0$ at all times. This last case agrees with the results presented in~\cite{PhysRevResearch.6.023213}, although the authors do not appear to make a distinction between even and odd values of $m$ for $\phi = m \pi$.

\subsection{Phase dependence in the non-Markovian regime}
\label{subsec:phase_dep}

\begin{figure}[h]
\centering
\includegraphics[width=\columnwidth]{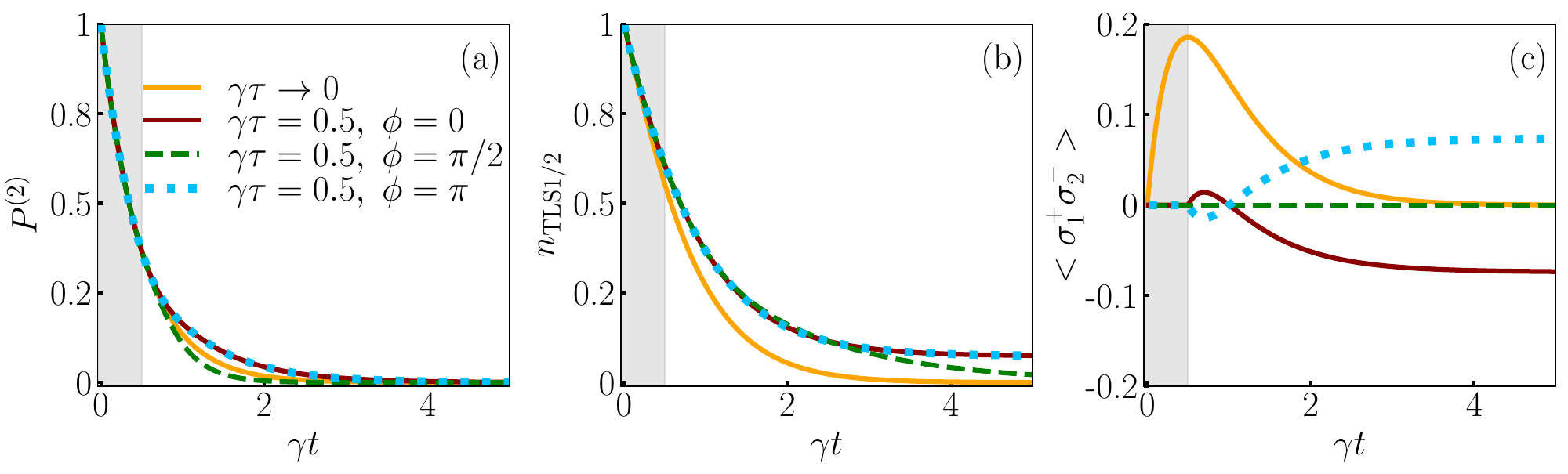}
\caption{Role of propagation phase on: (a) collective decay $P^{(2)}$, (b) individual populations $n_{\rm TLS1/2}$, and (c) atom-atom correlation $\braket{\sigma_1^+ \sigma_2^-}$ for a delay time of $\gamma \tau=0.5$ and three different phases $\phi=0$ (brown), $\phi=\pi/2$ (dashed green), and $\phi=\pi$ (dotted blue). The Markovian solution is plotted in orange for comparison.}
\label{fig:phases}
\end{figure}

In the regime of time-delayed feedback, the phase term $\phi$, defined earlier in Eq.~\eqref{eq:phase}, plays an important role in the time dynamics of the system, since it can control the onset of enhanced or suppressed atom decays. To appreciate the role of $\phi$, we have also investigated the impact on $P^{(2)}$, $n_{\rm TLS1/2}$, and $\braket{\sigma^+_1 \sigma^-_2}$ for a delay time of $\gamma\tau=0.5$ and compared with the Markovian result. 

First, in Fig.~\ref{fig:phases}(a), we show collective decay $P^{(2)}$ for three phases, $\phi=0$ (solid brown curve), $\phi=\pi/2$ (dashed green curve), and $\phi=\pi$ (dotted blue curve), and compare it with the results in the Markovian regime. As we already saw in Sec.~\ref{subsec:P2}, $P^{(2)}$ depends on $\tau$ in the non-Markovian regime and thus, it will also depend on $\phi$. Interestingly,  we observe how the collective decay is enhanced for a phase of $\phi=(m+ 1/2)\pi$, contrary to solutions with a single excitation, where the superradiant state is enhanced for a phase $\phi=2m\pi$~\cite{PhysRevLett.124.043603}.

Figure~\ref{fig:phases}(b) shows the individual populations $n_{\rm TLS 1/2}$, where the dynamics remain the same for both atoms, since they start in a symmetrical state and have symmetrical decay rates. For this observable, the fastest decay is observed in the Markovian solution, and $\phi=\pi/2$ does not decay faster than the other phases. The main difference that can be observed here is the population trapping for the phases, $\phi=\pi$ and $\phi=0$, which is not present in the case of $\phi=\pi/2$. 

Finally, in Fig.~\ref{fig:phases} (c), we show the time-dependent atom-atom correlations, $\braket{\sigma_1^+ \sigma_2^-}$, for the same phases. In this case, we see how atomic correlations are present even in the Markovian regime, except for $\phi=\pi/2$, which shows a perfect cancellation of the correlations between atoms. This again shows the dependence on the phase for the system evolutions, in agreement with the results in Ref.~\cite{PhysRevResearch.6.023213}.

\subsection{Additional quantum observables and non-Markovian effects}
\label{subsec:more_obs}

To better appreciate the role of non-Markovian effects on photons, we extend our study to include observables related to the explicit behavior of waveguide photons, and study both light-light and light-matter correlations.

\begin{figure}[h]
\centering
\includegraphics[width=0.95\columnwidth]{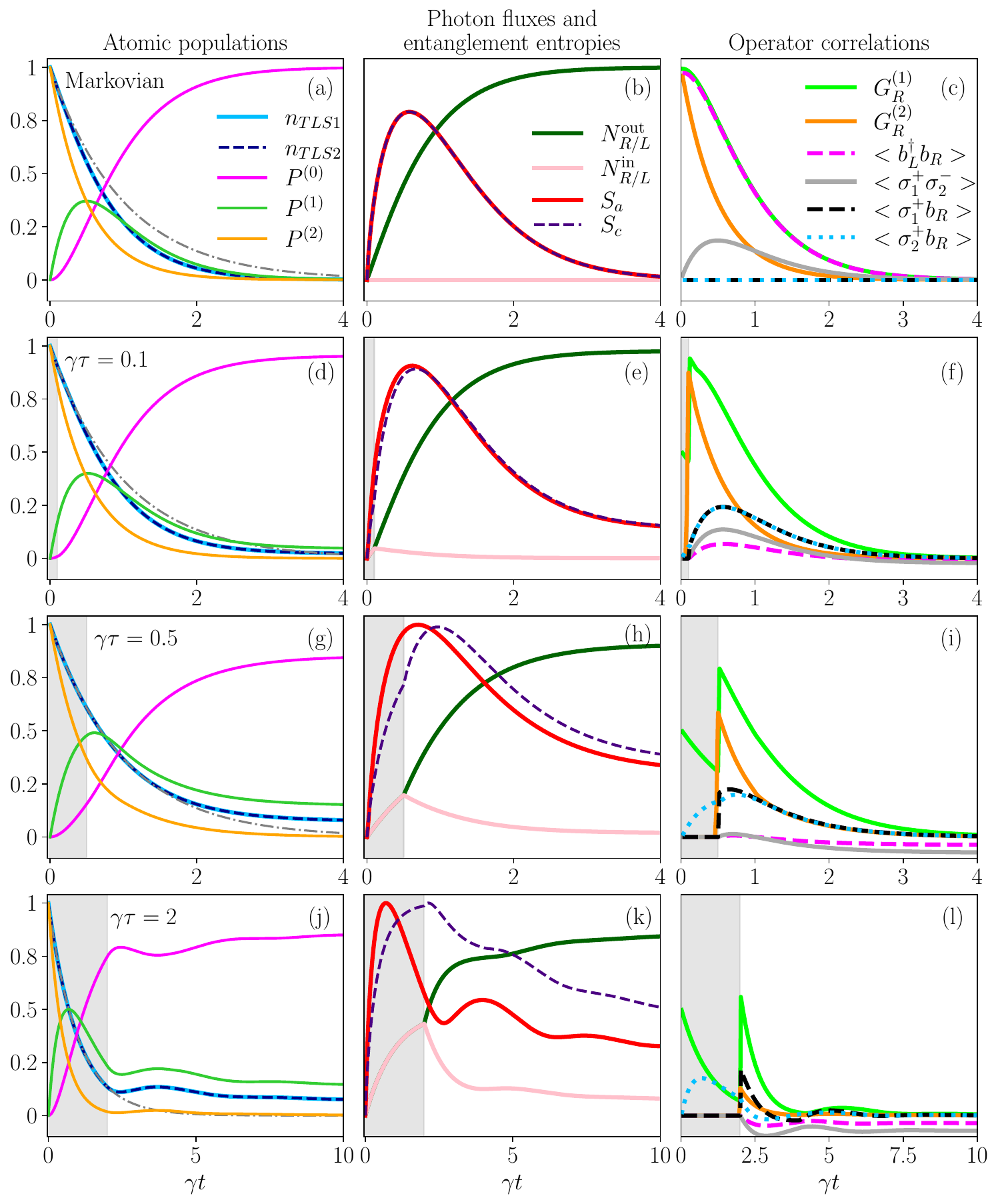}
\caption{Multi-panel plot summarizing various important observables for four different cases: Markovian regime (a,b,c);  $\gamma \tau=0.1$ (d,e,f); $\gamma \tau=0.5$ (g,h,i); and $\gamma \tau=2$ (j,k,l). (a,d,g,j) show the atomic population dynamics, (b,e,h,k) show photon fluxes and entanglement entropies, and (c,f,i,l) show correlations between operators in the system. }
\label{fig:TwoExc}
\end{figure}

In Fig.~\ref{fig:TwoExc}, we show a 
detailed summary
of populations, correlations, and entanglement entropies (introduced and defined below), first starting in the Markovian regime (a-c), and then with three different finite delay times [$\gamma \tau=0.1$ (d-f), $\gamma \tau=0.5$ (g-i) and $\gamma \tau=2$ (j-l)].
In the left column, the population dynamics are shown both for individual TLSs, and collective atomic excitations ($P^{(n)}$). Here, we observe, as before, the pronounced time-delayed feedback effects in the non-Markovian cases, where the atoms decay individually until the delay time, and then trapping of the TLS populations appears for longer times. Once again, with non-Markovian feedback, qualitatively different dynamics are observed for all $P^{(0)}$, $P^{(1)}$ and $P^{(2)}$. This is especially noticeable in the last two longer-delay cases 
[Fig.~\ref{fig:TwoExc} (g) and (j)], where we have longer delay times, $\gamma \tau=0.5$ and $\gamma \tau=2$, respectively. 

The center column figures show the integrated output flux of photons going to the left and right of the waveguide, $N^{\rm out}_{R/L}$ (dark green line), and the photon flux within the feedback loop, $N^{\rm in}_{R/L}$ (pale pink line), defined in Eqs.~\eqref{Ntotal} and~\eqref{Ntotaltau}. In this case, there is symmetry in the system, hence, the right and left moving photons will behave similarly. Here, we can see how, as the distance between the atoms increases (longer delay times), the output photon flux gets smaller at longer times. This is not only due to the trapped populations in the atoms, but also due to emerging probabilities of photons trapped in the feedback loop represented by $N^{\rm in}_{R/L}$. We stress that this observation is 
unique to the non-Markovian regime, since in the Markovian limit, the feedback loop is not considered.

In the (b,e,h,k) panels of Fig.~\ref{fig:TwoExc}, we also show the entanglement entropy, which is calculated in two different ways. First, we define the entanglement entropy between the atomic and the photonic part of the system, from~\cite{PhysRevLett.116.093601} 
\begin{equation}
S_a=-\sum_\beta \Lambda[\rm sys]^2_\beta  \log_2(\Lambda[sys]^2_\beta) , 
\end{equation}
where $\Lambda[\rm sys]_\beta$ are the Schmidt coefficients corresponding to the TLSs system bin, with $\beta$ representing the position of these coefficients, and the subscript `$a$' on $S_a$ stands for atomic. 
To ensure a maximum value of 1, this is normalized as $S_a= S_a/S_{\rm max}$ where $S_{\rm max} = n_{\rm qubits} \log_2 2$. Secondly, we define the entanglement entropy of the entire `circuit', $S_c$, which corresponds to the entropy between the `circuit' considered (the two TLSs plus the feedback or waveguide between them) and the rest of the waveguide,
\begin{equation}
S_c=-\sum_\beta \Lambda[\tau]^2_\beta \log_2(\Lambda[\tau]^2_\beta) , 
\end{equation}
where now $\Lambda[\tau]_\beta$ are the Schmidt coefficients of the feedback bin, and follow the same normalization $S_c= S_c/S_{\rm max}$.

In the Markovian regime,
both (entanglement) entropies introduced above behave in the same manner because there is no {\it feedback loop}, as shown in Fig.~\ref{fig:TwoExc}(b), and they simply increase to a certain value, and return to zero at sufficiently long times. However, as the feedback delay times increase [see Fig.~\ref{fig:TwoExc}(e,h,k)] so does the difference between these two observables, reflecting the increase in the probability of having photons trapped between the two TLSs. Moreover, the entropy values are no longer zero at long times, \textit{showing long term light-matter entanglement}. We can also observe oscillations in the entropy for long delay times ($\gamma \tau=2$), becoming clearer for $S_a$.

The right column panels in Fig.~\ref{fig:TwoExc} [(c,f,i,l)], show different types of quantum correlations. We first have the \textit{photonic correlations}, which are represented by the 
(one-time) first- and second-order quantum correlation functions $G_R^{(1)}(t)$ and $G_R^{(2)}(t)$ of the right output photons, and the correlation between right and left moving photons $\braket{b_L^\dagger b_R}(t)$. In the Markov limit, the first-order correlation function and $\braket{b_L^\dagger b_R}(t)$ are the same due to the symmetry of the system, but this is no longer the case when there are finite-delay feedback effects, where now the system behaves like two separated single-excited TLSs until the delay time ($\tau$), giving rise to different dynamics. This initial independent behavior is also reflected in the behavior of $G_R^{(2)}(t)$ which remains zero until $t=\tau$. 

The \textit{two-atom correlation} is also plotted on the right panels of Fig.~\ref{fig:TwoExc}(c,f,i,l). Once again, here we can see the difference between the Markovian regime, where there is an instantaneous correlation between the TLSs, and the non-Markovian one, where the correlation between atoms does not start until the delay time, and shows different dynamics depending on the atom distance, with negative correlation values for TLSs that are further apart.

Finally, we highlight that there are two more important observables shown in these figures, corresponding to the \textit{light-matter correlations} between the first TLS and the right moving photons, $\braket{\sigma_1^+ b_R}$, and the second TLS and the right moving photons, $\braket{\sigma_2^+ b_R}$. These two observables remain zero in the Markovian regime, where the light-matter entanglement is not present, and become different from zero when the feedback effects come into place, even when considering relatively short delay times like $\gamma \tau = 0.1$. This reinforces the fact that non-Markovian effects give rise to important light-matter entanglement not captured when considering certain approximations, and these can be measured from various correlation functions.

\section{Four coupled Qubits with two excitations}
\label{sec:results4}

Next we can extend our two-atom study by adding a new pair of two TLSs, i.e., four TLSs in total, and investigate how the dynamics differ from the previous cases. For this system, we study a delay time of $\gamma \tau=0.5$, where the delay effects are clearly visible but do not carry the oscillations of more extreme delay cases. 

In terms of the initial state, even if considering the same number of initial excitations (two atomic excitations), there are now several options for the initial quantum state of the system since the two new TLSs add more degrees of freedom. Here, we choose three example initial quantum states that we believe are especially interesting and relevant, but we remark that we can easily initialize the TLSs with any product state or entangled state containing two excitations, or in a more general case, containing up to four excitations. 

\textbf{Initial State A (one specific left atom excited
and one specific right atom excited):} Previously, for the case of two TLSs separated by a delay time $\tau$, initially excited, we had an initial atomic state as the following: $\ket{e_L,e_R}$, where $e_L$ represented the excited TLS on the left and $e_R$ the excited one on the right. Now we can add two additional TLSs, both starting in the ground state and see how these affect the collective dynamics of the system. In this case, we have the initial state, 
\begin{equation}
    \ket{\phi_0}_{A} = \ket{e_{1L},g_{2L},e_{1R},g_{2R}}.
\end{equation}

\textbf{Initial State B (two left atoms excited):} In the Markovian regime, when considering that the phase between TLSs is $\phi=0$, there is no difference for an initial product state with two excitations when choosing which two TLSs are excited. However, this is no longer the case when considering delay times. It is thus interesting to study the differences with an initial state where the two atomic excitations are on the left TLSs, since this case is only equivalent to the previous one (State A) in the Markovian limit. Here, we will have an initial state:  
\begin{equation}
    \ket{\phi_0}_{B} = \ket{e_{1L},e_{2L},g_{1R},g_{2R}}.
\end{equation}
 
\textbf{Initial State C (one unknown left atom excited and one unknown right atom excited):} In the previous examples, we assume that we know what precise TLSs are excited, but what happens if we only know that there is one excitation on the left and one on the right? The initial state will be a combination of all the possible initial states with two excitations, 
\begin{equation}
    \ket{\phi_0}_C = \frac{1}{2}\left( \ket{e_{1L},g_{2L},e_{1R},g_{2R}} + \ket{g_{1L},e_{2L},e_{1R},g_{2R}} +
\ket{g_{1L},e_{2L},g_{1R},e_{2R}}+
\ket{e_{1L},g_{2L},g_{1R},e_{2R}} \right).
\end{equation}

This case is particularly interesting, since if we rearrange the terms, we can also write the initial state as
\begin{equation}
    \ket{\phi_0}_C = \frac{1}{2} \left( \ket{e_{1L},g_{2L}} + \ket{g_{1L},e_{2L}} \right) \otimes \left( \ket{e_{1R},g_{2R}} + \ket{g_{1R},e_{2R}} \right)
\end{equation}
which corresponds to a product of a one excitation superradiant state of the left TLSs and a one excitation superradiant state with the TLSs on the right, and we can observe superradiant features in the collective population dynamics.

\begin{figure}[h]
\centering
\includegraphics[width=0.49\columnwidth]{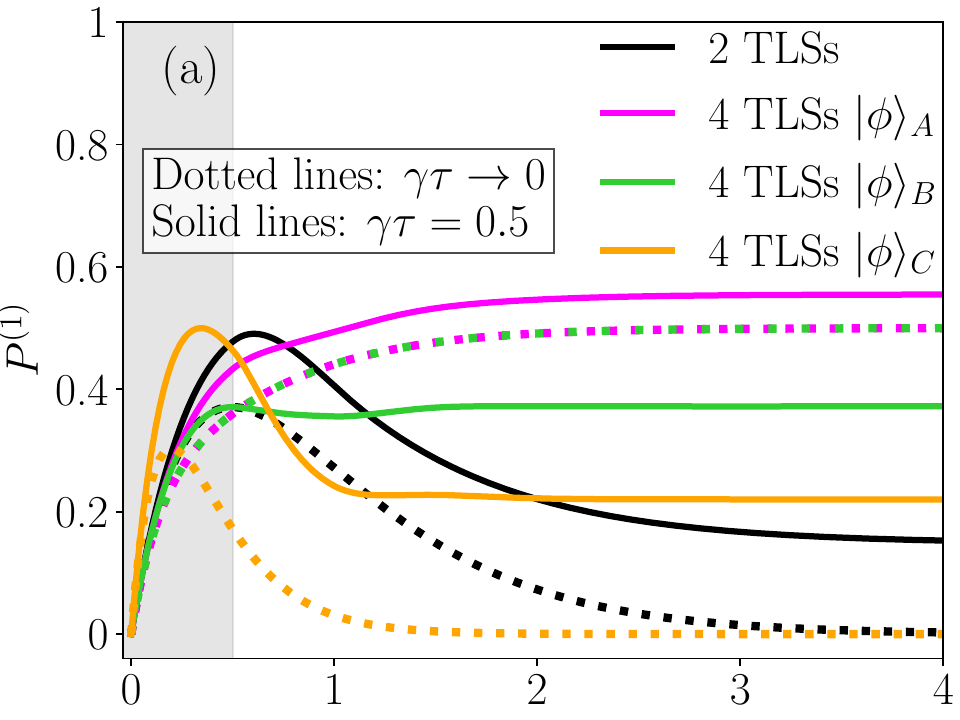}
\includegraphics[width=0.49\columnwidth]{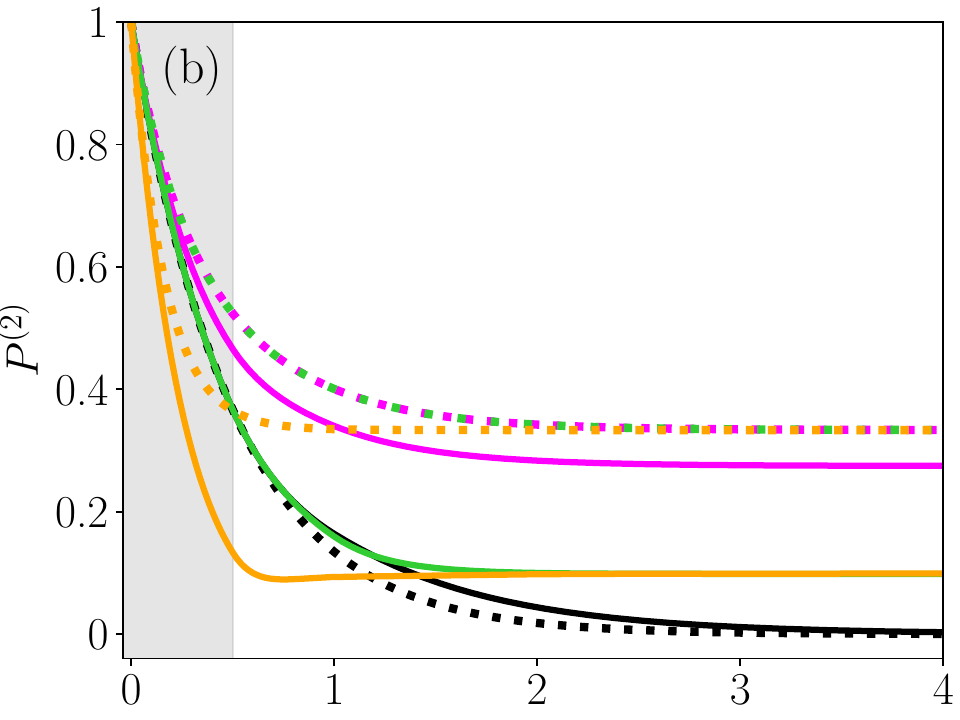}
\includegraphics[width=0.49\columnwidth]{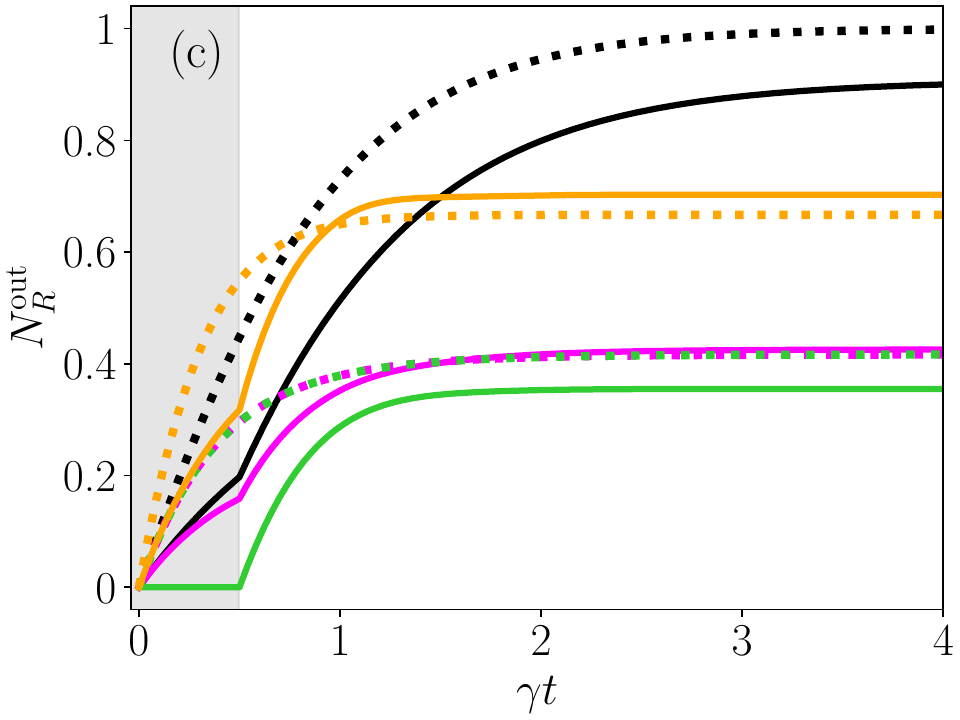}
\includegraphics[width=0.49\columnwidth]{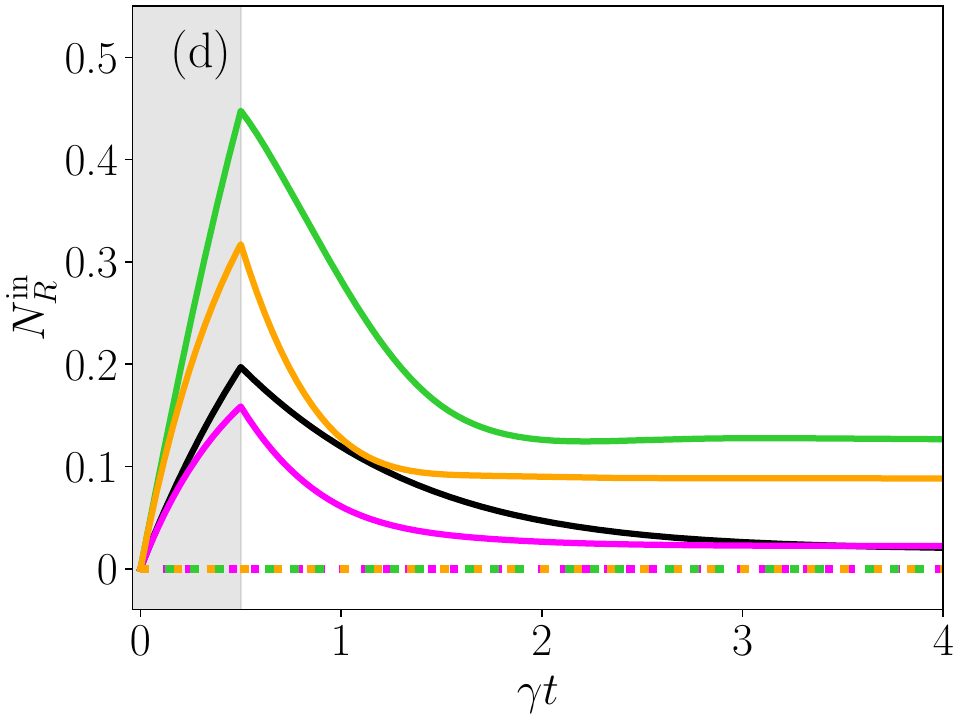}
\caption{(a) Probability of having one atomic excitation. (b) Probability of having two atomic excitations. (c) Integrated photon flux transmitted to the right. (d) Photon probability between the left and right TLS pairs. Magenta lines correspond to State A ($\ket{\phi}_A$), green lines to State B (($\ket{\phi}_B$)), orange lines to State C ($\ket{\phi}_C$), and black lines to the two excitation 2-TLSs solution previously studied. In all cases, the solid lines correspond to a delay time of $\gamma \tau=0.5$ and the dotted lines are in the Markovian limit. 
}
\label{fig:comp_cases}
\end{figure}

The time-dependent quantum dynamics of these three cases are 
summarized in Fig.~\ref{fig:comp_cases}, where State A ($\ket{\phi}_A$) is shown in magenta, State B ($\ket{\phi}_B$) in shown in green, and State C ($\ket{\phi}_C$) is shown in orange. For comparison, the results for two TLSs with two excitations are shown in black. Solid curves display non-Markovian results and dashed lines display Markovian results.

Several interesting features are observed. First, in Fig.~\ref{fig:comp_cases}(a), we can see how $\ket{\phi}_C$ behaves closer to the 2-TLSs case separated by a distance $d$ [see Fig.~\ref{fig:schem} (a)], both in the Markovian and non-Markovian cases, where in the Markovian limit the probability of one atomic excitation is zero in the long time limit, but it reaches a constant in the non-Markovian regime. For the initial state $\ket{\phi}_C$, the dynamics are faster with four TLSs, since this is a double superradiant state and, for the same delay time, it reaches a higher constant value. However, $\ket{\phi}_A$ and $\ket{\phi}_B$ have totally different dynamics from the two TLSs solution, both with and without considering the delay times. As expected, in the Markovian limit, the dynamics of $\ket{\phi}_A$ and $\ket{\phi}_B$ are the same but when feedback effects come into place they deviate, with an increasing $P^{(1)} (t)$ for $\ket{\phi}_A$ and decreasing values for $\ket{\phi}_B$.

In Fig.~\ref{fig:comp_cases}(b), we first see how with two TLSs, the probability of two atomic excitations is zero for long times (in both regimes), since photons escape into the external waveguide parts. However, when adding the extra two TLSs (four in total), the 2-excitation probability gets trapped in both the Markovian and non-Markovian regimes, as some waveguide photons are now trapped. This trapping in the Markovian regime is due to the fact that with four TLSs, even if we consider that their distance is small enough that we can disregard the feedback effects, we are still distinguishing the left and right pair of qubits. A fully superradiant state in which the two excitations were unknown within the four TLSs, would not have long lasting atomic excitations in the Markovian regime. 
In these cases, even though there are superradiant features, especially in State C, the distinction between pairs gives rise to atomic excitation trapping, even in the Markov limit.

Interestingly, in all the Markovian cases, the constant value at longer times is exactly the same value ($P^{(2)} \approx 0.33$), although it initially decays much faster if the initial state is $\ket{\phi}_C$, since again this is a combination of superradiant states. 
However, when considering the time-delayed feedback, we have three different collective decay rates. First, for $\ket{\phi}_A$, where we have two separated pairs of TLSs with one excitation in each side before the delay time, has the slowest rate, although this rate is still faster than the Markovian regime equivalent; and after the delay time both rates seem to become equal. Second, $\ket{\phi}_B$ has the same dynamics as the 2-TLSs case before ($\gamma \tau=0.5$), with a decay rate of $2\gamma$, since both excitations are on the left atoms, and after this time, the probability of having the two atomic excitations reaches a constant, smaller than in the previous case. Finally, $\ket{\phi}_C$ has a decay rate of $4\gamma$, having the fastest dynamics, and it reaches a trapped state after the delay time, $\tau$, similar to the previous one.   

Figures \ref{fig:comp_cases}(c) and (d) show the photon dynamics of the system, with (c) showing the right output photon flux $N_R^{\rm out}$ and (d) the photon probability in the waveguide part between the TLS pairs for the right moving photons $N_R^{\rm in}$, defined previously. From the simulations with four TLSs, $\ket{\phi}_A$ has a low transmitted photon flux and the lowest probability of having photons trapped between the TLS pairs, which agrees with the fact that both the one and two atomic excitation probabilities are higher than in the rest of cases. In constrast, $\ket{\phi}_B$ has fewer transmitted photons (with zero transmission until the photons reach the right pair of atoms when considering feedback) but it has the highest probability of having photons trapped between TLSs in the non-Markovian regime. The last case studied corresponding to $\ket{\phi}_C$, shows the largest transmission, with an even higher value when considering feedback effects, which, together with the highest $P^{(1)} (t \to \infty)$, and the lowest $P^{(2)} (t \to \infty)$, makes it a good example of a long term light-matter entangled state.

\section{Conclusions}
\label{sec:conclusions}

Using a powerful MPS approach, we have studied the effects of having finite delay times in the quantum dynamics of waveguide QED systems with multiple two-level atoms, with finite spatial separations, when they are prepared in superradiant states containing two atomic excitations. We have shown how the collective decay rates are 
significantly modified with larger distances between atoms and studied the impact on various light-matter correlations and the entanglement between the TLSs and the field propagating in the waveguide. We have also demonstrated the clear breakdown of assuming instantaneous coupling rates, leading to much richer collective effects than for a Markovian system.

Starting with two TLSs, we first studied the dependence of feedback effects on the double-excited state dynamics (superradiance regime), and on the emitter-emitter correlations. 
We showed how the superradiant coupling is {\it delayed} 
by the causality propagation time, manifesting in several non-Markovian decay regimes and long lived emitter-photon decay rates. Our theory fully includes the waveguide photon dynamics including the effects or roundtrip photons and population trapped states. 

We then extended our study to present a more detailed analysis of the quantum dynamics by calculating other observables, including entanglement entropy and light-matter correlations, and several of these correlations are unique only in the non-Markovian regime, which is accessible in quantum circuits. For example, photon-matter correlation effects are zero for the Markovian coupling regime (which assumes zero feedback time), but significant in the presence of finite delays. In addition, the entanglement entropy between the atomic and photonic part ($S_a$) and the entanglement entropy between the circuit considered and the rest of the waveguide ($S_c$), are equivalent in the Markovian regime, since the photonic region between TLSs is not considered, but we have shown how they qualitatively differ as delay times increase and the probability of photons in that region becomes larger.

Finally, to explore additional new waveguide QED superradiant regimes with more qubits,  we then added two more atoms to the waveguide, next to the original pair (thus having four qubits in total), and we again investigated their time dynamics when finite-delay feedback effects occur. Here, we have shown that, depending on the choice of our initial state, we can achieve different quantum dynamics, including long-lived light-matter entanglement.
Here, we found trapping of the collective doubly excited state with finite values at long times, in contrast to the 2 TLSs dynamics. We also observed trapping of the photons located between the TLSs. We have shown how these characteristics can be engineered with different initial states, making this a very rich system for creating light-matter entangled states.

Our general theory can easily include 
additional photons and qubits, allowing us to explore waveguide-mediated quantum coherence and entanglement generation in multi-atom waveguide QED systems, including cascaded chiral systems~\cite{PhysRevX.14.011020}.

\acknowledgements

This work is supported by the Natural Sciences and Engineering Research Council of Canada, the National Research Council of Canada, the Canadian Foundation for Innovation, and Queen's University, Canada.
S.H. and S.A.R. acknowledge RIKEN for support. F.N. is supported in part by the Japan Science and Technology Agency (JST)
[via the CREST Quantum Frontiers program Grant No. JPMJCR24I2,
the Quantum Leap Flagship Program (Q-LEAP), and the Moonshot R\&D Grant Number JPMJMS2061],
and the Office of Naval Research (ONR) Global (via Grant No. N62909-23-1-2074).
We also thank Alberto Del \'Angel Medina and Anton Frisk Kockum,
for valuable comments and insightful discussions.

\bibliography{references_all}

\end{document}